\documentclass[aps,superscriptaddress,twocolumn,]{revtex4}
\usepackage{bm}
\usepackage{epsfig}
\usepackage{times}
\usepackage{bbm}
\usepackage{amssymb}
\usepackage{amsmath}
\usepackage{natbib}
\usepackage{color}
\usepackage[latin1]{inputenc}
\usepackage[loose,nice]{units}       

\begin{document}
\title{Alternative gauge for the description of the light-matter interaction in a relativistic framework}
\author{Tor Kjellsson\footnote{tor.kjellsson@fysik.su.se}}
\affiliation{Department of Physics, Stockholm University, AlbaNova University Center, SE-106 91 Stockholm, Sweden}
\author{Morten F{\o}rre}
\affiliation{Department of Physics and Technology, University of Bergen, N-5007 Bergen, Norway}
\author{Aleksander Skjerlie Simonsen}
\affiliation{Department of Physics and Technology, University of Bergen, N-5007 Bergen, Norway}
\author{S{\o}lve Selst{\o}}
\affiliation{Faculty of Technology, Art and Design, Oslo and Akershus University College of Applied Sciences, NO-0130 Oslo, Norway}
\author{Eva Lindroth}
\affiliation{Department of Physics, Stockholm University, AlbaNova University Center, SE-106 91 Stockholm, Sweden}

\begin{abstract}
We present a generalized velocity gauge form of the relativistic laser-matter interaction. In comparison with the (equivalent) regular minimal coupling description, this new form of the light-matter interaction 
results in superior convergence properties for the numerical solution of the time-dependent Dirac equation. This applies both to the numerical treatment and, more importantly, to the multipole expansion of the laser field. The advantages of the alternative gauge is demonstrated in hydrogen by studies of the dynamics following the impact of superintense laser pulses of extreme ultraviolet wavelengths and sub-femtosecond duration.
\end{abstract}

\maketitle

\section{Introduction}
\label{Introduction}
With  high laser intensities,
available already  now or in the near future~\cite{Moore1999,DiChiara:08,Yumoto:13,Yoneda:14}, and the interesting possibilities then opening, discussed for example in Ref.~\cite{RevModPhysDiPiazza},
the description of the light-matter interaction in a relativistic framework is of growing importance. The ionization dynamics initiated with few-cycle laser pulses calls further for a time-dependent treatment.
Several attempts have consequently been made to solve the time-dependent Dirac equation (TDDE), see, e.g., Refs.~\cite{Selsto2009,Pindzola2010,Bauke2011,Vanne2012}, but it has been proven a hard task to explore the truly relativistic region while simultaneously accounting for the spatial dependence of the electromagnetic field and the full dimensionality of the problem. Recently, however,  a numerical study was  made where high orders of multipole interaction terms were successfully accounted for~\cite{kjellsson2017_TDDEvel}. Field intensities up to the strength where electrons are expected to reach quiver velocities, $v_\mathrm{quiv} \approx e E_0/ m\omega$, of around 20\% of the speed of light were treated, and emerging relativistic effects could be detected. Still, the study also  underlined some severe problems appearing when one is tackling the TDDE, concerning in particular the inclusion of magnetic effects.

When electrons are driven to high velocities by laser fields the magnetic part of the electromagnetic field inevitably becomes increasingly important. A qualitatively new effect  emerging is then the force imposed on the particle in the propagation direction of the light. Simulations  in the low- or medium-intensity regimes are usually made within the \textit{dipole approximation}, where the spatial dependence of the vector potential of the pulse is neglected completely. Since this approximation implies a neglect of all magnetic effects it is rather pointless in the high intensity regime~\cite{Reiss2000}. To understand the importance of different types of effects beyond the dipole approximation it is illustrative to look at the studies within this regime that have been done with the non-relativistic time-dependent Schr{\"o}dinger equation (TDSE). In that case the spatial dependence of the vector potential may conveniently be treated through a Taylor expansion~\cite{Forre2014,Simonsen-Forre::15} and the lowest order contribution has been shown to dominate the dynamics beyond the dipole completely -- at least up to intensities that drive the electron to velocities just above ten percent of the speed of light~\cite{Forre2014}. 
Surprisingly enough, as shown in Ref.~\cite{kjellsson2017_TDDEvel}, when the same approach is used with the TDDE, the lowest order spatial contribution from the Taylor expansion gives results that deviates significantly from the non-relativistic results already at modest intensities, far below the relativistic regime. This can be corrected by adding the next term in the expansion, but when the intensity is increased further the situation is repeated and one is forced to include also the following term and so on. This behavior can be analyzed and understood in the non-relativistic limit, as shown in Ref.~\cite{kjellsson2017_TDDEvel} and also discussed in Sec.~\ref{non-rel_limit} below. 
The problem stems from contributions that are known to cancel (approximately), but which enter  in different 
formal orders with respect to the Taylor expansion when it is applied to the Dirac equation.
Wherever the expansion is truncated, there will be unbalanced contributions which at some intensity will play a significant role. 
This imbalance is inherent to the four-component Dirac equation, and if the TDDE is to be solved for strong relativistic pulses the Taylor expansion approach in the \textit{regular} minimal coupling Hamiltonian quickly leads to an intractable problem.

Recently a \textit{generalized} velocity gauge form of the \textit{non-relativistic} light-matter interaction was presented~\cite{forre:2016,simonsen:forbidden:2016}. Within this gauge the dipole contribution is given as in velocity gauge, while the so called diamagnetic term disappears and instead new terms appear.
Of these, the leading order ones depend explicitly on the momentum in the direction of the laser propagation and the new gauge was consequently coined the {\em propagation gauge}.
It was further shown that it is possible to use a series of gauge transformations to 
successively remove all field-dependent terms that do not depend explicitly on the momentum.
The lowest order interaction within this gauge was further tested~\cite{forre:2016} and compared to simulations performed with the traditional minimal coupling Hamiltonian.  Impressive numerical advantages were then demonstrated. One manifestation of this was the evolution of the momentum expectation value along the direction of propagation of the light. In the minimal coupling description it showed a strong oscillating behavior, but in the new gauge it was replaced by a smooth curve that could be sampled with much larger time steps. Moreover, for a wave function expanded in spherical harmonics, the ionized wave packet could be described with considerably fewer angular momenta. In the following we will show that a corresponding gauge choice for the TDDE is even more advantageous. It requires just a single gauge transformation, it takes a simpler form, and it is a promising candidate for studying strong relativistic multipole interactions.

The paper is structured as follows: In the next section we outline the theoretical framework. Brief details on the implementation are provided in Sec.~\ref{Implementation}, while the results are presented and discussed in Sec.~\ref{Results}. Finally, we present our conclusions in Sec.~\ref{Conclusion}.
Atomic units are used throughout the text unless explicitly stated otherwise.

\section{Theory}
\label{Theory}

In a non-relativistic framework, the evolution of a wave packet representing a particle of mass $m$ and charge $-e$ in the scalar field $\varphi$ and vector potential ${\bf A}$ is governed by the TDSE,
\begin{equation}
\label{TDSE}
i \hbar \frac{\mathrm{d}}{\mathrm{d} t} \Psi_\mathrm{NR} = H_\mathrm{NR}(t) \Psi_\mathrm{NR},
\end{equation}
with the Hamiltonian
\begin{equation}
H_\mathrm{NR}(t)  = \left[ \frac{\mathbf{p}^2}{2m} -e\varphi + \frac{e}{m}
{\bf p} \cdot {\bf A} + \frac{e^2 A^2}{2m} \right] \ .
\label{Hamiltonian_TDSE}
\end{equation}
Here the potential ${\bf A}$ has been taken to fulfill the Coulomb gauge condition, $\nabla \cdot {\bf A} = 0$. Letting the electromagnetic pulse be defined in terms of the vector potential ${\bf A}$,
and assuming the field to be linearly polarized along the $z$ axis and propagating along the $x$ axis, 
the pulse may be written
\begin{equation}
\label{Adef}
{\bf A}(\eta) =  A(\eta) \hat{\bf z} = \frac{E_0}{\omega} f(\eta) \sin(\eta + \phi) \, \hat{\bf z} ,
\end{equation}
where $\eta = \omega t - k x$ and $k = \omega/c$. The envelope function is chosen to be sine squared:
\begin{equation}
\label{EnvelopeDef}
f(\eta) = \left\{ \begin{array}{lc} \sin^2 \left( \frac{\pi \eta}{\omega T} \right), & 0 < \eta < \omega T  \\ 0, & \text{otherwise} \end{array} \right. .
\end{equation}

In the dipole approximation, when the spatial dependence of the vector potential is neglected, the 
$A^2$ term can be removed by a gauge transformation and consequently does not affect the dynamics. On the other hand, this \textit{diamagnetic term} is known to give the leading contribution {\em beyond the dipole approximation} in the high-intensity limit. This has, e.g., been shown in Ref.~\cite{Forre2014}, 
where the spatial dependence of the vector potential was examined with the help of a Taylor series expansion:
\begin{align}
\label{taylor}
A(\eta) \approx \sum_{n=0}^{n_\mathrm{trunc}} \frac{1}{n!}
\frac{d^n A (\eta)}{d \eta^n} \bigg\rvert_{\eta=\omega t}  \left(- \frac{\omega x}{c} \right)^n .
\end{align}
In Ref.~\cite{Forre2014} $n_\mathrm{trunc} \leq 2$ was considered. A point worth noticing here is that the diamagnetic term is second order in $\mathbf{A}$, and thus an expansion of $A$ to a particular order in $(\omega x/c)^n$ does not imply an expansion of the Hamiltonian to the same order. The contributions to the Hamiltonian with, for example, $n=2$ come from the square of the $n=1$ term in Eq.~(\ref{taylor}) {\em and} from the cross term between the $n=0$ and $n=2$ terms. Furthermore, there should be considerable cancellations between these terms; their sum oscillates with twice the frequency of the light while each of them have a constant sign. Large cancellations were indeed found in Ref.~\cite{Forre2014}, and it was concluded that it is decisive to include all terms that contribute to the {\em Hamiltonian} to a given  
order. When the corresponding time-dependent Dirac equation is solved using the minimal coupling Hamiltonian~\cite{kjellsson2017_TDDEvel}, the diamagnetic term, which is then only implicitly included, causes severe convergence problems in terms of the multipoles of the external field. This is connected to an effective blocking of the aforementioned cancellations as will be clear in the following.

Turning now to the TDDE, the first step will be to consider the minimal coupling Dirac Hamiltonian;
\begin{equation}
\label{Hamiltonian}
H(t) = c \boldsymbol{\alpha} \cdot \left[ {\bf p} +  e{\bf A}(\eta) \right] -e \varphi(r) \mathbbm{1}_4 + m c^2 \beta,
\end{equation}
and
\begin{equation}
\label{AlphaMat}
\boldsymbol{\alpha}
= \left( \begin{array}{cc} 0 &
\boldsymbol{\sigma}
\\
\boldsymbol{\sigma}
&  0 \end{array}\right).
\end{equation}
 As usual,  $\boldsymbol{\sigma}$ is given by the Pauli matrices, and
\begin{equation}
\label{BetaMat}
\beta = \left( \begin{array}{cc} \mathbbm{1}_2 & 0 \\  0 &  -\mathbbm{1}_2 \end{array} \right).
\end{equation}
We set out to solve the TDDE:
\begin{equation}
\label{TDDE}
i \hbar \frac{\mathrm{d}}{\mathrm{d} t} \tilde{\Psi} = H(t)\tilde{\Psi},
\end{equation}
where the four-component wave function $\tilde{\Psi}$ can be written as
\begin{equation}
\label{EigenStatesOfH}
\tilde{\Psi}({\bf r},t)
=
\left( \begin{array}{c} \tilde{\Psi}_F({\bf r},t) \\ \tilde{\Psi}_G({\bf r},t)\end{array} \right),
\end{equation}
with $\tilde{\Psi}_F$ and $\tilde{\Psi}_G$ being two-component spinors, often called the large and small component, respectively. The potential $\varphi(r)$ is for the present purposes simply the Coulomb potential from a point nucleus, i.e., we neglect retardation effects in the electron-nucleus interaction and take the nuclear mass to be infinite.

\subsection{The non-relativistic limit of the relativistic minimal coupling Hamiltonian}
\label{non-rel_limit}
In order to understand the origin of the problems encountered with the TDDE expressed in terms of the orignal minimal coupling Hamiltonian, it is important to study its non-relativistic limit. Since we are aiming for a solution to the TDDE which describes a positive energy state, we may write
\begin{equation}
\tilde{\Psi}({\bf r},t) = \Psi({\bf r},t) e^{-i mc^2 t}.
\end{equation}
Eq.~(\ref{TDDE}) can then be rewritten as:
\begin{equation}
\label{TDDEnew}
i \hbar \frac{\mathrm{d}}{\mathrm{d} t} \Psi({\bf r},t) = \left( H(t) -mc^2 \right)\Psi({\bf r},t).
\end{equation}
Using the form of the wave function given in Eq.~(\ref{EigenStatesOfH}) we can write Eq.~(\ref{TDDEnew})
as two coupled differential equations:
\begin{align}
-e \varphi \Psi_F +c \boldsymbol{\sigma} \cdot \left( \mathbf{p} + 
e \mathbf{A}  \right) \, \Psi_G = i\hbar \frac{\mathrm{d} \Psi_F}{\mathrm{d} t} \nonumber \\
c \boldsymbol{\sigma}\cdot \left( \mathbf{p} +  e \mathbf{A} \right)   \Psi_F
+ \left( -e \varphi - 2mc^2 \right) \, \Psi_G = i\hbar \frac{\mathrm{d} \Psi_G}{\mathrm{d} t}.
\label{twokomp}
\end{align}
If only the dominating terms on the second line is retained (i.e. assuming that the mass-energy term is large both compared to the potential energy, $2mc^2 \gg e\varphi$, and to the time variation of the small component) it is possible to write the small component as:
\begin{align}
\label{smallcomp}
\Psi_G \approx  \frac{1}{2mc} \boldsymbol{\sigma} \cdot \left( \mathbf{p} + e \mathbf{A} \right)   \Psi_F.
\end{align}
When inserting this into the first line of Eq.~(\ref{twokomp}) we get
\begin{align}
-e \varphi \Psi_F +\frac{1}{2m} \left( \boldsymbol{\sigma} \cdot \left( \mathbf{p} + e \mathbf{A} \right)\right)^2 \, \Psi_F = i\hbar \frac{\mathrm{d} \Psi_F}{\mathrm{d} t},
\label{step1}
\end{align}
and with some operator algebra, detailed in Ref.~\cite{kjellsson2017_TDDEvel}, this expression can be rewritten as
\begin{align}
\left( \frac{\mathbf{p}^2}{2m} -e \varphi
+ \frac{e}{m} \mathbf{p}\cdot \mathbf{A}
+\frac{e^2 A^2}{2m}
+ \frac{e \hbar}{2 m}  \boldsymbol{\sigma}
\cdot \mathbf{B} \right) \Psi_F
= i\hbar \frac{\mathrm{d} \Psi_F}{\mathrm{d} t},
\label{step2}
\end{align}
where $\mathbf{B} = \nabla \times \mathbf{A}$. Apart from the spin-dependent term, the operators on the left-hand side are the same as those that appeared in Eq.~(\ref{Hamiltonian_TDSE}). In particular, we note the diamagnetic contribution, which apparently is implicitly included in the TDDE through the coupling between the small and large component of the wave function. Hence the advice from Ref.~\cite{Forre2014} regarding the consistent inclusion of the $x^n$ terms from the Taylor expansion of $\mathbf{A}$ is not easy to follow for the TDDE. Since the Dirac equation is linear in the vector potential, truncation after a particular $n$ in Eq.~(\ref{taylor}) will result in an implicitly included diamagnetic term that contains the $x^{2 n}$ contributions from the square of the $x^n$ term, but not the cross terms between higher- and 
lower-order terms that are also $x^{2 n}$ contributions. As a consequence, the solution of the TDDE with only the lowest order spatial correction ($n=1$) to the vector potential generally gives meaningless results, as demonstrated in Ref.~\cite{kjellsson2017_TDDEvel}. Furthermore, the convergence of the dynamics with respect to $n_\mathrm{trunc}$ in Eq.~(\ref{taylor}) was shown to be very slow once the laser pulse parameters started to enter the relativistic regime. In the case where the electron was accelerated to a quiver velocity of $v_{\mathrm{quiv}} \sim 0.2c$, a fifth order expansion was necessary for converged results. It is natural to assume that an even higher order expansion would be necessary further into the relativistic regime, and with each additional term $x^n$ in Eq.~(\ref{taylor}), the computational demand quickly turns this into an intractable problem. It is thus highly relevant to instead seek an alternative route less prone to grow so complex when the dynamics become increasingly relativistic.

\subsection{The relativistic propagation gauge}
Since it is the actual way the diamagnetic contribution resurfaces in the TDDE that causes the convergence problems, it might be possible to find an alternative form where it is easier to balance the terms included in Eq.~(\ref{taylor}). We are for instance free to make a gauge transformation to change the scalar field and vector potential as;
\begin{align}
\mathbf{A} \rightarrow \mathbf{A} + \nabla \zeta \nonumber \\
\varphi \rightarrow \varphi - \frac{\partial \zeta}{\partial t},
\end{align}
which will yield a transformed Hamiltonian:
\begin{align}
 H = c \boldsymbol{\alpha} \cdot \left[ {\bf p} +  e{\bf A}(\eta)  + e  \nabla \zeta \right]  +   \left[ e  \frac{\partial \zeta}{\partial t} - e \varphi(r) \right] \mathbbm{1}_4  + m c^2 \beta.
\end{align}
In Ref.~\cite{forre:2016} it was shown that by choosing a gauge that followed the classical electron momentum in the direction of the light propagation, $p_k$, the diamagnetic term in the Schr{\"o}dinger equation could be removed and replaced by operators that showed superior convergence properties. In the relativistic case, as shown in Refs.~\cite{PhysRevD.1.2738,PhysRevA.54.4383}, a free classical particle that is initially at rest will acquire the momentum
\begin{align}
\label{pk}
p_k(\eta) =  \frac{mc}{2} 
\left( \frac{{e A}(\eta) }{mc}\right)^{2},
\end{align}
when exposed to the electromagnetic field given by $\bf A(\eta)$. 
It is natural to assume that a suitable gauge can be found if $\zeta$ is defined using Eq.~(\ref{pk}), but we
start by defining it with an additional operator, $\aleph(\eta)$, that remains to be determined:
\begin{align}
\zeta \left( \eta \right) & =  - \frac{mc^2}{e \omega}  \int_{-\infty}^{\eta}
d\eta' 
\frac{1}{2}\left( \frac{{e A}(\eta') }{mc}\right)^{2} 
\aleph\left( \eta' \right).
\end{align}
The introduction of $\aleph$ is related to the distinction between the relativistic and the non-relativistic version of the gauge transformation leading to the propagation gauge formulation~\cite{forre:2016,simonsen:forbidden:2016}. 
We will return to its specific form in the following.

With the vector potential polarized along the $z$ axis and the field propagating along the $x$ axis we obtain
\begin{align}
\label{newterm}
e  \nabla \zeta = - \frac{mc^2}{\omega}   \hat{\bf x} \, \frac{\partial \eta}{\partial x}
\frac{d \;}{d \eta}
\int_{-\infty}^{\eta}
d\eta' \frac{1}{2} 
\left( \frac{e A(\eta') }{mc}\right)^{2} 
\aleph\left( \eta'\right) 
=
\nonumber \\
 + \hat{\bf x} \, k \frac{mc^2}{\omega} \frac{1}{2}
\left( \frac{e  A(\eta)}{mc}\right)^{2}  
\aleph\left( \eta\right)
=   
\hat{\bf x} \, mc
\frac{1}{2}\left(\frac{e A(\eta)}{mc}  \right)^{2}
\aleph\left( \eta\right) \, ,
\end{align}
where $k=\omega/c$ has been used in the last step. 

Equation~(\ref{newterm}) is a vector operator in the propagation direction of the field.
Further, with
\begin{align}
 e  \frac{\partial \zeta}{\partial t} =
 - \frac{m c^2 }{2}
\left( \frac{{e A}(\eta) }{mc}\right)^{2}  
\aleph\left( \eta\right),
\end{align}
we may now write down the {\em propagation gauge} Dirac Hamiltonian, $H_{\rm PG}$:
\begin{align}
\label{HPG}
H_{\rm PG} = c \boldsymbol{\alpha} \cdot \left[ {\bf p} +  e \mathbf{A}(\eta) \right]
 -e \varphi(r) \mathbbm{1}_4  + m c^2 \beta  \nonumber \\
 + \frac{e^2 A^2(\eta)}{2 m}  
 \aleph\left( \eta\right)
 \left(  \alpha_x
-  \mathbbm{1}_4 \right),
\end{align}
where the first line is just the minimal coupling Dirac Hamiltonian from Eq.~(\ref{Hamiltonian}). The second line, on the other hand, displays one operator proportional to $\alpha_x$, the relativistic velocity operator in the direction of the propagation of the light, and one {\em counter term}.
As we will see, this counter term cancels the implicit diamagnetic term contributed by the first line when the equation is examined in the non-relativistic limit.

 \subsection{The non-relativistic limit of the relativistic propagation gauge}
Starting again from Eq.~(\ref{TDDEnew}) but now adding the new terms from the second line in Eq.~(\ref{HPG}), we will instead of Eq.~(\ref{twokomp}) find
\begin{align}
- \left( \varphi + \frac{e^2 A^2 }{2m} \aleph  \right)\Psi_F +\nonumber \\
\left( c \boldsymbol{\sigma} \cdot \left( \mathbf{p} + e \mathbf{A} \right) +
\sigma_x   \frac{e^2 A^2 }{2m} \aleph\right) \, \Psi_G = i\hbar \frac{\mathrm{d} \Psi_F}{\mathrm{d} t} \label{twokompPG1} \\
\left( c \boldsymbol{\sigma} \cdot \left(  \mathbf{p} + e \mathbf{A} \right) +
\sigma_x \frac{e^2 A^2 }{2m} \aleph  \right)  \Psi_F \nonumber \\
- \left( e \varphi +  2mc^2 + \frac{e^2 A^2 }{2m} \aleph \right) \, \Psi_G = i\hbar \frac{\mathrm{d} \Psi_G}{\mathrm{d} t} .
\label{twokompPG}
\end{align}
Following the derivation preceding Eq.~(\ref{smallcomp}), and assuming in addition that $2mc^2$ dominates also over $(e^2 A^2/2m) \aleph$, we obtain a new approximate relation between the large and small component:
\begin{align}
\label{smallcompPG}
\Psi_G \approx  \frac{1}{2mc} \left( \boldsymbol{\sigma} \cdot \left(   \mathbf{p} + e \mathbf{A}\right)  +
\sigma_x   \frac{e^2 A^2 }{2mc} \aleph  \right) \Psi_F.
\end{align}
Inserting this expression for $\Psi_G$ into Eq.~(\ref{twokompPG1}) we find the propagation gauge Hamiltonian in the non-relativistic limit (cf. the expression for the minimal coupling Hamiltonian on the left-hand side of Eq.~(\ref{step2}));
\begin{align}
\label{HNRPG}
H_{\rm PG}^{\rm NR} = \frac{\mathbf{p}^2}{2m}  + \frac{e}{m} \mathbf{p} \cdot \mathbf{A}  + 
\frac{e^2 A^2}{2m} 
-e \phi  +  \frac{e \hbar}{2 m}  \boldsymbol{\sigma} \cdot \mathbf{B} \nonumber \\
+ \frac{1}{2 m c} \left\{\frac{e^2 A^2 }{2m} \aleph,p_x\right\}
- \frac{e^2 A^2 }{2m}\left(\aleph  - \aleph^2 \frac{e^2 A^2 }{4 m^2 c^2}\right),
\end{align}
where $ \left\{a,b\right\}$ denotes an anticommutator.
In addition to the original terms in Eq.~(\ref{step2}), two new terms have appeared on the last line of Eq.~(\ref{HNRPG}).
It is evident that if we put $\aleph=1$ 
the diamagnetic term is cancelled. 
However, another possibility is to require
\begin{align}
\label{thisiszero}
\frac{e^2 A^2}{2m} - \frac{e^2 A^2 }{2m}\left(\aleph  - \aleph^2 \frac{e^2 A^2 }{4 m^2 c^2}\right) = 0
\end{align}
and thus get rid also of the term proportional to $A^4$. If Eq.~(\ref{thisiszero}) is regarded as the defining equation for $\aleph$, we can readily write down its expression as
\begin{align}
\label{closedaleph}
\aleph = \frac{1 - \sqrt{1-\left(\frac{e A}{mc}\right)^2}}
{\frac{1}{2}\left(\frac{e A}{mc}\right)^2} ~.
\end{align}
It is clear from Eq.~(\ref{closedaleph}) that its range of validity is restricted to the region where
\begin{align}
\left(\frac{e A}{mc}\right)^2 < 1~,
\end{align}
which is consistent with the approximation made to obtain Eq.~(\ref{smallcompPG}). In this case we may also expand Eq.~(\ref{closedaleph}) and find
\begin{align}
\label{aleph_exp}
\aleph = 1 +\frac{1}{4}\left(\frac{e\mathbf{A}}{mc}\right)^2 +
\frac{1}{8}\left(\frac{e\mathbf{A}}{mc}\right)^4 +
\frac{5}{64}\left(\frac{e\mathbf{A}}{mc}\right)^6 + \ldots ~,
\end{align}
which in fact is the series that was found in Refs.~\cite{forre:2016,simonsen:forbidden:2016}, i.e.,
\begin{align}
\label{AlephExpansion}
&\aleph =  \sum_{j=0}^{\infty} 2 a_{j+1} \left(\frac{e\mathbf{A}}{mc}\right)^{2j}
 \\ \nonumber
{\rm with} \quad & a_j  =  \frac{(2j)!}{4^j (2j-1) (j!)^2} ~ .
\end{align}

With $\aleph$ defined this way we may write Eq.~(\ref{HNRPG}) as
\begin{align}
H_{\rm PG}^{\rm NR}  = \frac{\mathbf{p}^2}{2m}  + \frac{e \mathbf{p} \cdot \mathbf{A}}{m} -e \phi  +  \frac{e \hbar}{2 m}  \boldsymbol{\sigma} \cdot \mathbf{B}
\nonumber \\
+ \frac{1}{2 m c} \left\{\frac{e^2 A^2 }{2m} \aleph,p_x\right\},
\end{align}
which, apart from the spin-dependent term, is identical to the propagation gauge Hamiltonian obtained directly from the TDSE in Refs.~\cite{forre:2016,simonsen:forbidden:2016}.

\subsection{The long-wavelength approximation}
\label{Theory_LWA}
While a vector potential without spatial dependence does not introduce any magnetic interaction in the ordinary minimal coupling Dirac Hamiltonian, Eq.~(\ref{Hamiltonian}), a purely time-dependent ${\bf A}$ {\it does} provide an additional dynamical term in Eq.~(\ref{HPG}); the term proportional to $\alpha_x$.
Again this is in agreement with the findings in Refs.~\cite{forre:2016,simonsen:forbidden:2016};
in the propagation gauge the radiation pressure is accounted for through a velocity gauge-like operator acting along the propagation direction of the laser in spite of a spatially independent $\mathbf{A}$.
The effective Dirac Hamiltonian in this {\em long-wavelength approximation} (LWA) is given by
\begin{align}
\label{LWA}
H_{\rm LWA} = c \boldsymbol{\alpha} \cdot \left[ {\bf p} +  e \mathbf{A}(\omega t) \right]
 -e \varphi(r) \mathbbm{1}_4  + m c^2 \beta
\nonumber \\
+   \frac{e^2 A^2(\omega t)}{2 m} \aleph(\omega t)
  \alpha_x,
\end{align}
where the terms that lack spatial dependence altogether have been removed since they do not affect the dynamics. In Sec.~\ref{Results} we will show that the Hamiltonian Eq.~(\ref{LWA}) can account fully for the dominating effects beyond the dipole approximation for a wide range of electromagnetic pulses.
It gives in fact excellent agreement with the much more demanding fifth order expansion 
of the Hamiltonian Eq.~(\ref{Hamiltonian}), as applied in Ref.~\cite{kjellsson2017_TDDEvel}.

In principle we are free to choose either $\aleph =1$, to follow the relativistic momentum in the direction of the propagation of the laser light, or as given in Eq.~(\ref{closedaleph}) to allow for a more straight forward comparison with the non-relativistic treatment. For high enough fields there will of course be differences for any non-exact implementation, as will be demonstrated in Sec.~\ref{Results}. Lastly, although the properties of the LWA-Hamiltonian Eq.~(\ref{LWA}) are very promising, we want to emphasize that a practical implementation of Eq.~(\ref{HPG}) is by no means restricted to the LWA-approximation. It is indeed possible to go further and introduce spatial dependence in $\mathbf{A}$, which should become important for large enough field strengths $E_0$ and/or in the limit of very high laser frequencies. For illustrative purposes we present a first order beyond the LWA Hamiltonian in the next section and later demonstrate that it gives negligible contributions for the laser pulses considered in this article, which accelerate the electron to quiver velocities $v_{\rm quiv}$ up to about $0.2c$.

\subsection{Beyond the long-wavelength approximation}
We introduce a spatial dependence in $\mathbf{A}$ by using $n_{\rm{trunc}}=1$ in Eq.~(\ref{taylor}). Then, using $\aleph=1$, the first order beyond the long-wavelength approximation (BYLWA1) Hamiltonian can be written as
\begin{align}
\label{BYLWA1}
H_{\rm BYLWA1} = c \boldsymbol{\alpha} \cdot \left[ {\bf p} +  e \mathbf{A}(\omega t) \right] - 
e \alpha_z x \omega A'(\omega t) \nonumber \\
 -e \varphi(r) \mathbbm{1}_4  + m c^2 \beta
 \nonumber \\ +
 \frac{e^2}{2m}
  \left(  \frac{x \omega}{c} 2 A(\omega t) A'(\omega t) - \frac{x^2 \omega^2}{c^2}  
  (A'(\omega t))^2 \right) \, \mathbbm{1}_4
 \nonumber \\
 + \,  \frac{e^2}{2m} \left(  A^2(\omega t)  -  \frac{x \omega}{c} 
 2 A(\omega t)A'(\omega t)\right) \, \alpha_x \, . \nonumber \\
\end{align}
It may seem odd that the two terms proportional to $A^2$ in Eq.~(\ref{HPG}) have been expanded differently. 
However, according to the discussion in Sec.~\ref{non-rel_limit}, this is indeed the proper way of expanding 
the field-dependent terms as this minimizes the problem with inconsistent terms appearing in the corresponding non-relativistic Hamiltonian. Note also that terms lacking spatial dependence altogether have been removed from Eq.~(\ref{BYLWA1}). 

In the continuation, we will demonstrate that the Hamiltonian $H_{\rm LWA}$ in Eq.~(\ref{LWA}) provides practically all dynamics for fields penetrating into the relativistic region, simply by comparing its results with the corresponding results obtained with $H_{\rm BYLWA1}$ as defined above. For clarity we emphasize that $\aleph=1$ has been used in both Hamiltonians for a just comparison. Before presenting our results we briefly describe our numerical implementation.

\section{Implementation}
\label{Implementation}
We expand the wave function in eigenstates of the time-independent Hamiltonian, i.e., Eq.~({\ref{Hamiltonian}) without $\mathbf{A}$, giving
\begin{equation}
\label{Expansion}
\Psi(t) = \sum_{n,j,m,\kappa} c_{n,j,m,\kappa}\left( t \right) \psi_{n,j,m,\kappa}({\bf r}),
\end{equation}
with
\begin{equation}
\label{EigenStatesOfH0}
\psi_{n,j,m,\kappa}({\bf r})
=
\left( \begin{array}{c} F_{n,j,m,\kappa}({\bf r}) \\ G_{n,j,m,\kappa}({\bf r})\end{array} \right),
\end{equation}
where
\begin{equation}
\label{EigenStatesOfH0_radial}
\left( \begin{array}{c} F_{n,j,m,\kappa}({\bf r}) \\ G_{n,j,m,\kappa}({\bf r})\end{array} \right)
 = \frac{1}{r} \left( \begin{array}{c} P_{n,\kappa}(r) X_{\kappa,j,m}(\Omega) \\ i Q_{n,\kappa}(r) X_{\textrm{-}\kappa,j,m}(\Omega) \end{array} \right).
\end{equation}
Here $\kappa=l$ for $j=l-1/2$ and $\kappa=-(l+1)$ for $j=l+1/2$. $X_{\kappa,j,m}$ represents the spin-angular part which has the analytical form
\begin{equation} 
\label{AngularPart}
\mathit{X}_{\kappa, j, m} = \sum_{m_s,m_l} \langle l_{\kappa},m_l ;s,m_s | j,m \rangle {Y}^{l_{\kappa}}_{m_l} (\theta, \phi) \chi_{m_s},
\end{equation}
where ${Y}^{l_{\kappa}}_{m_l} (\theta, \phi)$ is a spherical harmonic and $\chi_{m_s}$ is an eigenspinor. The radial components $P_{n,\kappa}(r)$ and $Q_{n,\kappa}(r)$ are expanded in B-spline functions~\cite{Boor1978};
\begin{equation} \label{P_Q_Bspl} P_{n,\kappa}(r) = \sum_{i} a_i B_i^{k_1}(r), \quad
Q_{n,\kappa}(r) = \sum_{j} b_j B_j^{k_2}(r).
\end{equation}
Just as in Ref.~\cite{kjellsson2017_TDDEvel} we use B-spline functions of orders $k_1=7$ and $k_2 = 8$ for the small and large components, respectively. As has been shown by Froese Fischer and Zatsarinny~\cite{Charlotte2009}, the use of different $k$ for the two components effectively removes the so called spurious states which are known to appear when the Dirac equation is solved within a finite basis set. We use a linear knot sequence with 500 B-spline functions for the large component and 501 for the small component up to $R_\mathrm{max} = 150$~a.u.. To avoid unphysical reflections at the box boundary we have used a complex absorbing potential starting from $r = 110$~a.u.. We include all spin-orbitals with angular momenta up to a certain $l_\mathrm{max}$ (as defined for the large component) and keep all the associated magnetic quantum numbers $m$, cf.~Eqs.~(\ref{EigenStatesOfH0_radial}, \ref{AngularPart}). To speed up the propagation without compromising the results, high energy components have been filtered out from the basis.

In Sec.~\ref{Results} we present converged data for the energy distribution, the expectation value of the momentum operator along the pulse propagation direction and, finally, the total ionization yield 
from the hydrogen ground state exposed to a $15$ cycle, $95$~eV ($\omega=3.5$~a.u.) laser field of intensity $7\times 10^{19}$~W/cm$^2$ ($E_0=45$~eV). For the two former quantities, converged data were obtained with $l_{\rm max}=30$ for the propagation gauge LWA, cf.~Eq~(\ref{LWA}), corresponding to 1,902,594 states and about $2.32 \times 10^{10}$ non-zero matrix elements. In order to arrive at the same result with the minimal coupling Hamiltonian, Eq.~(\ref{Hamiltonian}), it was necessary to include $l_{\rm max}=50$ for the case with $n_\mathrm{trunc}=5$ in Eq.~(\ref{taylor}). From now on we will refer to this level of approximation  as {\em fifth order beyond dipole} (BYD5). The BYD5 simulation required 5,125,954 states and about $1.64 \times 10^{12}$ non-zero matrix elements, i.e., roughly 70 times more than our converged propagation gauge simulations.

For the ionization yield, which was systematically investigated for both lower and higher values of $E_0$, convergence was always achieved with $l_{\rm max} = 40$ for both $H_{\rm LWA}$, Eq.~(\ref{LWA}), and $H_{\rm BYLWA1}$, Eq.~(\ref{BYLWA1}). For further details on the implementation, such as how interaction matrix elements are computed and which numerical schemes that are applied, readers are referred to Ref.~\cite{kjellsson2017_TDDEvel}.

\section{Results}
\label{Results}
We will first present results for the following scenario: A hydrogen atom with the electron initially prepared in the ground state is exposed to a laser pulse, defined in Eqs.~(\ref{Adef} - \ref{EnvelopeDef}), with the parameters
\begin{equation}
\label{laser_params}
\begin{array}{lc} E_0 = 45.0~\mbox{a.u.},  \ \ \omega = 3.5~\mbox{a.u.},  \ \ \phi = 0, \\
\text{and} \ \ T = N_c \tfrac{2 \pi}{\omega}~ \mbox{a.u.} \ \ \text{with} \ \ N_c = 15. \end{array}
\end{equation}
The pulse parameters are such that the electron's quiver velocity is expected to reach about $v_{\mathrm{quiv}} \sim 0.1c$ and have been chosen primarily to demonstrate the convergence property of the relativistic propagation gauge -- not to reveal relativistic effects per se. To show the convergence properties, the lowest order interaction in the propagation gauge, LWA, cf.~Eq.~(\ref{LWA}), has been compared to the minimal coupling Hamiltonians ranging from BYD1 to BYD5, that is with $n_\mathrm{trunc}=1-5$ in Eq.~(\ref{taylor}). Figure~\ref{dPdE_1} shows a comparison of the energy distribution of the ionized electron after interaction with the pulse. A somewhat typical convergence pattern for the minimal coupling simulations can be seen, where each successive interaction type pushes the distribution to either side of the fully converged result.

\begin{figure}[h]
\includegraphics[width=0.50\textwidth]{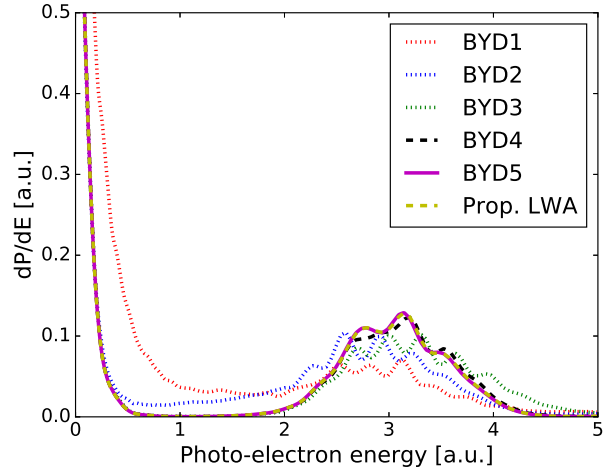}
\caption{Kinetic energy spectrum of the emitted photoelectron following the laser-assisted 
ionization of the hydrogenic ground state, with the laser field described in Eq.~(\ref{laser_params}).
Results obtained with the minimal coupling Hamiltonian Eq.~(\ref{Hamiltonian}) with the laser field 
treated at five different level of approximations, i.e., in increasing complexity from BYD1 to BYD5, as well
as the result  obtained with the propagation gauge Hamiltonian in lowest order, i.e., the LWA in~Eq.~(\ref{LWA}), are shown for comparison. The minimal coupling results are starting to converge with BYD4, and with BYD5 there is a good agreement with the propagation gauge result.
\label{dPdE_1}}
\end{figure}

Figure~\ref{log10_dPdE} also shows the energy distribution but now only for the minimal coupling BYD5 result 
and the propagation gauge LWA result. The energy grid has been extended to include the three first ionization peaks and a logarithmic scale is used to better resolve the data. The coincidence between the LWA result in the propagation gauge, which only involves purely time-dependent fields, and the result using a fifth order Taylor expansion of the vector field within the minimal coupling formulation Eq.~(\ref{Hamiltonian}), is evident. 
\begin{figure}[h]
\includegraphics[width=0.50\textwidth]{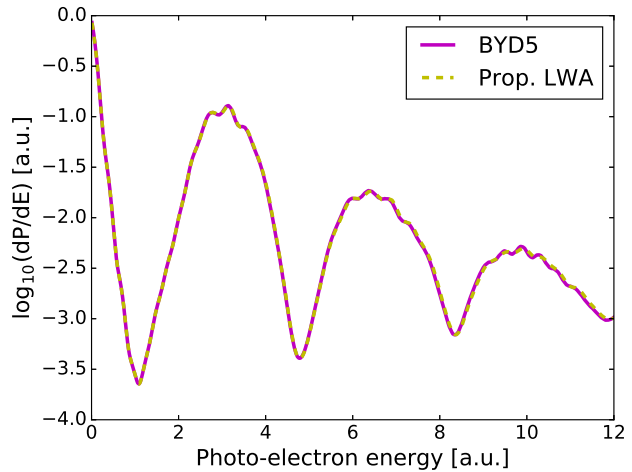}
\caption{ As Fig.~\ref{dPdE_1}, but a comparison between the converged BYD5 minimal coupling result and the corresponding propagation gauge result obtained within the LWA. The energy grid has been extended and a logarithmic scale is used for higher resolution.}
\label{log10_dPdE}
\end{figure}

In the non-relativistic version of the propagation gauge~\cite{forre:2016} an important demonstration of its computational advantages was the smooth evolution of the expectation value of the momentum in the propagation direction of the laser field. The same behaviour is found also for the corresponding relativistic results in Fig.~\ref{px_expval}, where violent oscillations seen using the minimal coupling Hamiltonian are transformed into a smooth development in the propagation gauge. Interestingly, to the naked eye, $\langle p_x \rangle$ seems to be converged already at BYD3 while the probability distribution clearly requires at least BYD5, as seen in Fig.~\ref{dPdE_1}.

\begin{figure}[h]
\includegraphics[width=0.50\textwidth]{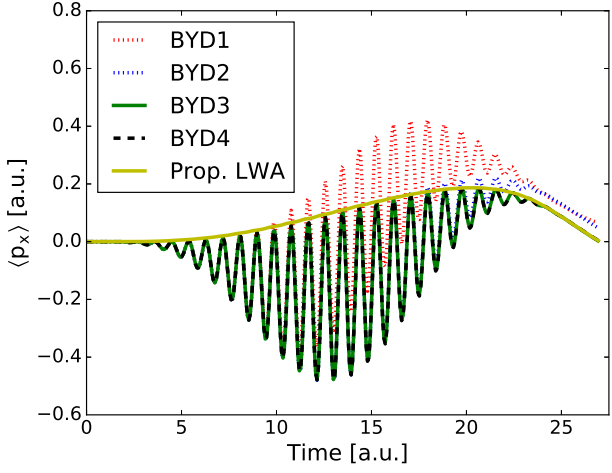}
\caption{Expectation value of the momentum in the propagation direction of the pulse, as 
obtained with BYD1 to BYD4 as well as the propagation gauge LWA, and 
for the laser field given in Eq.~(\ref{laser_params}). The total pulse duration is $\sim26.9$~a.u..}
\label{px_expval}
\end{figure}

A comparison with non-relativistic simulations is also in order. As mentioned, the chosen pulse parameters result in an expected quiver velocity of $v_{\mathrm{quiv}} \sim 0.1c$ and only small relativistic corrections, if any, are expected. Figure~\ref{log10_dPdE_comp} shows a comparison of the relativistic and non-relativistic probability distributions, as obtained by solving both the TDSE in the propagation gauge LWA, cf.~Ref.~\cite{forre:2016}, and the TDDE with the propagation gauge LWA Hamiltonain, Eq.~(\ref{LWA}), 
using an equivalent basis set in both cases. Indeed, there are no relativistic effects displayed in Fig.~\ref{log10_dPdE_comp}, and one only expects these to appear at even higher intensities.

\begin{figure}[h]
\includegraphics[width=0.50\textwidth]{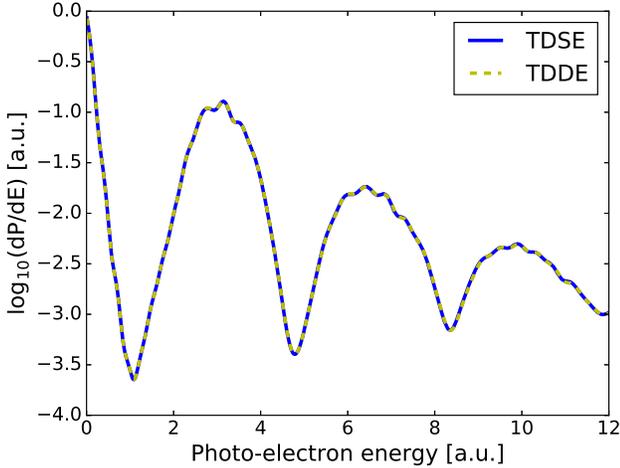}
\caption{As Fig.~\ref{dPdE_1}, but a comparison of the relativistic and non-relativistic results, as obtained by the TDDE and TDSE, respectively.}
\label{log10_dPdE_comp}
\end{figure}%

In order to search for possible relativistic effects we now systematically increase the field strength up to about $E_0 = 100$ a.u., corresponding to $I \sim 3.5 \times 10^{20} \mbox{ W/cm}^2$ and $v_{\mathrm{quiv}} \sim 0.2c$. Figure~\ref{survival} shows the resulting ionization yield as a function of the electric field strength. The minimal coupling results from Ref.~\cite{kjellsson2017_TDDEvel} (TDDE BYD5), the 
relativistic propagation gauge results obtained both within the LWA (Eq.~(\ref{LWA})) and beyond (Eq.~(\ref{BYLWA1})), as well as the corresponding TDSE result, are shown for comparison. Again, the relativistic corrections seem to be very small. Nevertheless, a tiny relativistic shift manifested as a decrease in the ionization yield, is displayed as the quiver velocity approaches $v_{\rm quiv} \sim 0.2c$. Furthermore, Fig.~\ref{survival} shows that the favorable behavior of the LWA propagation gauge Hamiltonian persists over a wide range of intensities, up to the onset of the relativistic regime. 

\begin{figure}[h]
\includegraphics[width=0.50\textwidth]{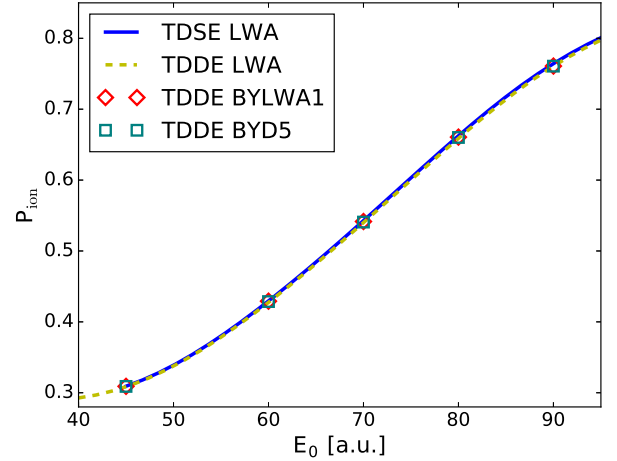}
\caption{Ionization yield of a hydrogen atom irradiated by a laser pulse with the parameters from Eq.~(\ref{laser_params}) and varying peak electric field strength $E_0$. The relativistic propagation gauge results, obtained both within the LWA and beyond, i.e., Eqs.~(\ref{LWA}) and~(\ref{BYLWA1}), both agree with the minimum coupling BYD5 results from Ref.~\cite{kjellsson2017_TDDEvel}. A comparison with the corresponding TDSE LWA result reveals a small relativistic correction, manifested as a decrease in the ionization yield as the quiver velocity approaches $v_{\rm quiv} \sim 0.2c$. $E_0 =90$~a.u. corresponds to a peak intensity of $2.8 \times 10^{20}$~W/cm$^2$.}
\label{survival}
\end{figure}

Finally, going even further into the relativistic regime the important question of how to properly incorporate the full spatial dependence of the field in the propagation gauge Dirac Hamiltonian needs to be addressed. In an exact calculation, the choice of $\aleph$, be it simply $\aleph=1$ or as in Eq.~(\ref{closedaleph}) or, equivalently, Eq.~(\ref{AlephExpansion}), does of course not matter. However, with a truncated representation of the field, cf.~Eq.~(\ref{taylor}), this choice may be of 
crucial importance -- and increasingly so for increasing field strengths. This dependence has been investigated and the first results are shown in Fig.~\ref{LWA_aleph}. As in Fig.~\ref{survival}, ionization probabilities are shown  for both the TDSE and the TDDE, but here for field strengths $E_0$ in the interval $80-105$~a.u. and only within the LWA. In going from  $\aleph$ as defined in Eq.~(\ref{closedaleph}) to
simply choosing $\aleph=1$, i.e., truncation at the first term in Eq.~(\ref{AlephExpansion}),
a small shift downwards is introduced -- consistently both in the relativistic and non-relativistic treatments, respectively. Based on the present results, it is still unclear which of the choices for $\aleph$ represent the best approximation, nor where the LWA approximation breaks down. However, the correspondence between the gauge transformation and the relativistic momentum in the propagation direction for a corresponding free, classical electron moving in the field, cf.~Eq.~(\ref{pk}), suggests that simply $\aleph=1$ should be the best choice. Although we here leave these open questions for future research, simply due to the computational complexity of the problem, it should be noted that they could all be studied within the current computational framework.

\begin{figure}[h]
\includegraphics[width=0.50\textwidth]{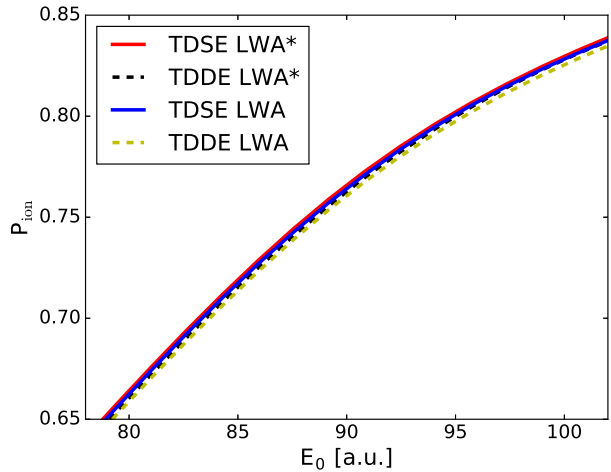}
\caption{As Fig.~\ref{survival}, but with the ionization yield obtained within the LWA for different choices of the truncation in Eq.~(\ref{AlephExpansion}). The "*" in the legend indicates simulations obtained when truncating the expansion after 10 terms, while $\aleph=1$ is used in the other two graphs (see text for details). Both results obtained with the TDSE and the TDDE are shown for comparison. In both the relativistic and the non-relativistic cases a visible shift with respect to the truncation level appears from $E_0 \sim 60$ a.u. (see Fig.~\ref{survival}), and becomes pronounced from $E_0 \sim 80$ a.u. $E_0 =100$~a.u. corresponds to a peak intensity of $3.5 \times 10^{20}$~W/cm$^2$.}
\label{LWA_aleph}
\end{figure}

\section{Conclusion}
\label{Conclusion}
We have presented a generalized velocity gauge form of the relativistic light-matter interaction and demonstrated its superior convergence properties compared to the regular minimum coupling Hamiltonian. As in the non-relativistic case, the alternative relativistic gauge relaxes the requirement on the maximum angular momentum needed during the time propagation. However, the major advantage goes even beyond that. While the usual minimal coupling formulation is numerically tough for high-intensity fields treated in a non-relativistic framework, it constitutes an intractable problem in the relativistic case due to inherent imbalance in the Dirac equation. The propagation gauge to a large extent removes this imbalance and opens up for calculations on atoms subjected to electromagnetic pulses in the truly relativistic regime.

\section*{Acknowledgments}
Simulations have been performed partly on resources provided by the Swedish National Infrastructure for Computing (SNIC) at TRIOLITH and KEBNEKAISE and partly on resources provided by the Norwegian Metacenter for Computational Science (UNINETT Sigma2) at HEXAGON (Account number NN9417K). Financial support by the Swedish Research Council (VR), Grant No. 2016-03789, is gratefully acknowledged. We also acknowledge support for our collaboration through the {\it Nordic Institute for Theoretical Physics} (Nordita) and from the research group {\it Mathematical Modelling} at Oslo and Akershus University College of Applied Sciences.



\end{document}